\documentclass[twocolumn]{svjour3}          
\usepackage{graphicx,amsmath,amssymb,color}
 \usepackage{mathptmx}   
\usepackage[T1]{fontenc}
\usepackage[utf8]{inputenc}

\begin{document}

\title{Comment on ``Two-dimensional third- and fifth-order nonlinear evolution equations for shallow water waves with surface tension'' [Nonlinear Dyn, doi:10.1007/s11071-017-3938-7]}
\titlerunning{Comment on M.~Fokou et al.\ Nonlinear Dynamics (2018) 91:1177-1189.}  

\author{ Piotr Rozmej  \and  Anna Karczewska}

\institute{P. Rozmej \at Faculty of Physics and Astronomy,
University of Zielona G\'ora, Szafrana 4a, 65-246 Zielona G\'ora, Poland \\
              \email{P.Rozmej@if.uz.zgora.pl}           
           \and
           A. Karczewska \at Faculty of Mathematics, Computer Science and Econometrics, University of Zielona G\'ora, Szafrana 4a, 65-246 Zielona G\'ora, Poland\\
              \email{A.Karczewska@wmie.uz.zgora.pl}           
}

\date{Received: \today / Accepted: }

\maketitle

\begin{abstract}
The authors of the paper ``Two-dimensional third- and fifth-order nonlinear evolution equations for shallow water waves with surface tension'' \cite{Fok} claim that they derived the equation which generalizes the KdV equation to two space dimensions both in first and second order in small parameters. Moreover, they claim to obtain soliton solution to the derived first order (2+1)-dimensional  equation. The equation has been obtained by applying the perturbation me\-thod~\cite{burde} for small parameters of the same order. The results, if correct, would be significant. In this comment, it is shown that the derivation presented in \cite{Fok} is inconsistent because it violates fundamental properties of the velocity potential. Therefore, the results, particularly the new evolution equation and the dynamics that it describes, bear no relation to the problem under consideration.

 \end{abstract}

\section{Outline of the metod}

The authors of \cite{Fok} consider the model of inviscid and incompressible fluid which motion is irrotational in a container with a flat bottom. 
In dimensional variables, the set of hydrodynamical equations has the following form:
\begin{eqnarray}  \label{g1}
\phi_{xx} + \phi_{yy} + \phi_{zz}&=& 0, \quad \mbox{in~the~whole~volume}, \\ \label{g2}
\phi_z - (\eta_x \phi_x + \eta_y \phi_y +\eta_t) &=& 0, \quad \mbox{at~the~surface},  \\ \label{g3}
\phi_t + \frac{1}{2}(\phi_x^2+\phi_y^2+\phi_z^2) +g\eta &=& 0, \quad \mbox{at~the~surface}, \\ \label{g4}
\phi_z  &=& 0\; \quad \mbox{at~the~bottom}.
\end{eqnarray}
Here $\phi(x,y,z,t)$ denotes the velocity potential, $\eta(x,y,t)$ denotes the surface function and $g$ is the gravitational acceleration. Indexes denote partial derivatives, i.e.\ $\phi_{xx}\equiv \frac{\partial^2 \phi}{\partial x^2}$, and so on. The authors take into account surface tension terms, as well.
In this note, we neglect these terms since their presence or absence does not change the source of errors made in \cite{Fok}.

Next, the authors introduce a standard scaling to dimensionless variables (different in $x$- and $y$-direction)
\begin{align} \label{bezw}
   \tilde{x}= & x/L, \quad \tilde{y}= y/y_{0},\quad \tilde{z}= z/h_{0}, \quad \tilde{t}= t/t_{0},  \\
\quad \tilde{\eta}= &\eta/A,\quad \tilde{\phi}= \phi /(L\frac{A}{h_{0}}\sqrt{gh_{0}}), 
\end{align}
where $A$ is the amplitude of surface distortions from equilibrium shape (flat surface), $h_{0}$ is average fluid depth, $L$ is the average wavelength (in $x$-direction), and $y_{0}$ is a wavelength in $y$-direction. In general,  $y_{0}$ should be of the same order as $L$, but not necessarily equal. Notation $t_{0}=L/\sqrt{gh_{0}}$ is not explained in the paper.

Then the set (\ref{g1})-(\ref{g4}) takes in scaled variables the following form (tildas are now dropped):
\begin{eqnarray}  \label{G1}
\beta \phi_{xx} + \gamma\phi_{yy} + \phi_{zz}&=& 0,  \\ \label{G2}
 \eta_t +\alpha (\eta_x\phi_x+\frac{\gamma}{\beta}\eta_y\phi_y)-\frac{1}{\beta}\phi_z &=& 0,~ \mbox{for}~ z=1+\alpha\eta,\\ \label{G3}
\phi_t + \frac{1}{2}\alpha \left(\phi_x^2+\frac{\gamma}{\beta}\phi_y^2+\frac{1}{\beta}\phi_z^2\right) + \eta &=& 0,~ \mbox{for}~ z=1+\alpha\eta, \\ \label{G4}
\phi_z  &=& 0\;\,~  \mbox{for}~z=0.
\end{eqnarray}
As usual, small parameters are defined as follows: $$\alpha=A/h_{0}, \quad \beta=(h_{0}/l)^{2}\quad  \mbox{and}\quad \gamma=(h_{0}/y_{0})^{2}.$$

\section{Details of calculations in \cite{Fok} limited to first order}

Next, the authors assume $\gamma=\beta$ and write {\bf erroneous formula} \cite[Eq.~(11)]{Fok} for the velocity potential
\begin{equation} \label{F11}
\phi =\sum_{m=0}^{N} \sum_{n=0}^{N} \frac{(-1)^{m+n}\beta^{m+n} z^{2(m+n)} }{(2m)!(2n)!} \frac{\partial^{2(m+n)}}{\partial x^{2m}\partial y^{2n}} f(x,y,t).
\end{equation}

In the following, the authors limit their considerations to the Boussinesq equations up to second order in small parameters. Then it is enough to use  the explicit form of the potential (\ref{F11}) up to third order (due to terms $\frac {1}{\beta}\phi_z$ in (\ref{G2})  and $\frac {1}{\beta}\phi_z^2$ in (\ref{G3}) the velocity potential should be valid up to one order higher than the Boussinesq equations)
\begin{align} \label{F11s}
\phi = f & -\frac{\beta\,z^{2}}{2} \left(f_{xx}+f_{yy}\right)+ \frac{\beta^{2}z^{4}}{24}\left(f_{xxxx}+6f_{xxyy}+f_{yyyy}\right) \nonumber \\ & - \frac{\beta^{3}z^{6}}{720}\left(f_{6x}+15f_{4x2y}+15f_{2x4y}+f_{6y}\right).
\end{align}
\emph{The fact that formulas (\ref{F11})-(\ref{F11s}) are wrong is easy to check  by a direct substitution to the Laplace equation (\ref{G1}).}
With the above form of the velocity potential, the authors obtained the set of Boussinesq's equations in the form (here surface tension is neglected and only terms of first order are retained)
\begin{align} \label{g1F}
\eta_{t} + f_{xx} +f_{yy} & +  \alpha  \left(f_{x}\eta_{x}+f_{y}\eta_{y}+\eta(f_{2x}+f_{2y})\right) \nonumber\\ & -\frac{1}{6}  \beta
   \left(f_{xxxx} +6 f_{xxyy} +f_{yyyy} \right)  = 0, \\
\label{g2F}
\eta_{x} + f_{xt} & + \alpha  \left(f_{x} f_{xx} + f_{y}f_{xy} \right) 
 -\frac{1}{2} \beta \left(f_{xyyt} +f_{xxxt}  \right)  =0.
\end{align}
Next, the authors insert the velocity potential into equations (\ref{G2}) and (\ref{G3}) retaining terms up to second order in $\alpha,\varepsilon=\beta$. 
They introduce the following notation 
\begin{equation}\label{uv} u = f_{x}, \qquad v= f_{y}.\end{equation}
With the above notation Boussinesq's equations (\ref{g1F})-(\ref{g2F}) can be formulated as \cite[Eq.~(12)]{Fok}-\cite[Eq.~(13)]{Fok}, that is as
\begin{align} \label{g1f}
\eta_{t} + u_{x} +v_{y} &+  \alpha  \left(u\eta_{x}+v\eta_{y}+\eta(u_{x}+v_{y})\right) \\ & \nonumber
 -\frac{1}{6} \varepsilon
   \left(u_{xxx} +3 u_{xyy} + 3 v_{xxy} +v_{yyy} \right)  = 0, \\ 
\label{g2f}
\eta_{x} + u_{t} & + \alpha  \left(u u_{x} + v v_{x} \right) 
 -\frac{1}{2} \varepsilon \left(v_{xyt} +u_{xxt}  \right)  =0.
\end{align}

In the following, the authors apply the perturbative approach described in detail by Burde and Sergyeyev \cite{burde} and next extended in \cite{KRcnsns} to more complicated cases. In this me\-thod, one begins from zeroth-order solutions, then uses their properties in the calculation of corrections of the first order, and so on.
In zeroth order eqs.\ \cite[Eq.~(12)]{Fok}-\cite[Eq.~(13)]{Fok} reduce to
\begin{equation} \label{b0ord}
\eta_{t} + u_{x} +v_{y} =0, \qquad \mbox{and} \qquad \eta_{x} + u_{t} =0,
\end{equation}
which have solutions $u=\eta$, $v=a(x,t)$ with $\eta_{t}=-\eta_{x},~ u_{t}=-u_{x}$. The authors of \cite{Fok} take constant $v=a$, which is a particular solution to $v_{y}=0$.

Looking for the solution in first-order approximation, one introduces corrections of the first order to the equations (\ref{b0ord}) and requires that the Boussinesq's equations become {\bf compatible} in this order (what means that these two equations become equivalent). The authors look for first-order solutions in the form
\begin{eqnarray} \label{c1aF}
u &=& \eta +\alpha\, B(x,y,t) +\varepsilon\, C(x,y,t), \\ \label{c1bF}
v &=&  a\qquad \mbox{(constant )}.
\end{eqnarray}
Inserting first order corrections (\ref{c1aF})-(\ref{c1bF}) into (\ref{g1f})-(\ref{g2f}), retaining only terms up to first order, and using properties $B_{t}=-B_{x}, ~C_{t}=-C_{x}$ the authors obtained Here, we cite the formulas from \cite{Fok}. More detailed derivation, with the correct velocity potential formula, is presented in section \ref{s3}. 
These equations, after integration give the formulas \cite[Eq.~(23)]{Fok}-\cite[Eq.~(24)]{Fok}. 
These formulas read as
\begin{equation} \label{Accor}
B=-\frac{1}{4}\eta^{2}-\frac{1}{2} a \int\eta_{y}\,dy, \qquad C= \frac{1}{4}\eta_{yy}+\frac{1}{3}\eta_{xx}.
\end{equation}
So, in first order approximation $u$ becomes (in \cite[Eq.~(25)]{Fok} terms at $\alpha$ and $\varepsilon$ are incorrectly positioned)
$$ u = \eta +\alpha \left(-\frac{1}{4}\eta^{2}-\frac{1}{2} a \int\eta_{y}\,dy \right) +\varepsilon\left( \frac{1}{4}\eta_{yy}+\frac{1}{3}\eta_{xx}
\right).$$

Insertion of $u$, given by the above expression 
into  \cite[Eq.~(12)]{Fok} and \cite[Eq.~(13)]{Fok} leads in both cases to the same equation \cite[Eq.~(26)]{Fok}, that is
\begin{equation} \label{Fok26}
\eta_{t} +\eta_{x}+\alpha\left(\frac{3}{2}\eta\eta_{x}+\frac{1}{2}a\eta_{y}\right)+\varepsilon \left(\frac{1}{6}\eta_{xxx}-\frac{1}{4}\eta_{xyy}\right) =0. \end{equation} 
In other words, equation \cite[Eq.~(25)]{Fok}  supply compatibility of Boussinesq's equations and therefore may supply the correct (2+1)-dimensional evolution equation in the first order. On this basis, the authors extend the derivations to second order in small parameters and claim to obtain the (2+1)-dimensional fifth-order evolution equation \cite[Eq.~(32)]{Fok}. They also claim to find the analytic solution to the equation  \cite[Eq.~(26)]{Fok} and discuss its properties.

We do not comment here the derivation of second-order equation nor the authors' solutions to the first order equation because already first order wave equation {\bf is incorrect}. The details of the argumentation that the equation (\ref{Fok26}), that is, \cite[Eq.~(26)]{Fok}, is wrong are presented in section \ref{s4}.

\section{Details of correct calculations in first order perturbation approach} \label{s3}

The correct formula for the velocity potential fulfilling (\ref{g1}) has the following form 
\begin{align} \label{Szer1} \phi(x,z,t) & =\sum_{m=0}^\infty \frac{(-1)^m}{(2m)!} z^{2m}\, (\beta\partial_{xx}+\gamma\partial_{yy})^m f(x,y,t). \end{align} 
For $\gamma=\beta$,  the explicit form of this velocity potential up to third order in small parameters reads as 
\begin{align} \label{pot8} \phi = f & -\frac{\beta\,z^2 }{2}(f_{xx}+f_{xy}) + \frac{\beta^2 z^4 }{24}( f_{xxxx}+2 f_{xxyy}+f_{yyyy}) \nonumber \\ & - \frac{\beta^3 z^6}{720} (f_{6x}+3f_{4x2y}+3f_{2x4y}+f_{6y}). \end{align}
The correct formula for the velocity potential (\ref{Szer1}) and (\ref{pot8}) differs from that used by the authors ((\ref{F11}) and (\ref{F11s})) by values of coefficients in front of mixed $xy$-derivatives.

With the correct formula (\ref{pot8}), one obtains the following Boussinesq's equations. From~(\ref{G2}), limiting to first order, one gets
\begin{align} \label{g1a}
\eta_{t} + f_{xx} +f_{yy} & + \alpha  \left(f_{x}\eta_{x}+f_{y}\eta_{y}+\eta(f_{2x}+f_{2y})\right) \nonumber  \\ & -\frac{1}{6}  \epsilon 
   \left(f_{xxxx} +2 f_{xxyy} +f_{yyyy} \right)  = 0. \end{align}
The correct result from (\ref{G3}) in first order (after differentiation over $x$) is 
\begin{align} \label{g2a}
\eta_{x} + f_{xt} & + \alpha  \left(f_{y}f_{xy}  +f_{x} f_{2x} \right) 
 -\frac{1}{2} \epsilon \left(f_{xyyt} +f_{xxxt}  \right) 
  =0.
\end{align}
In variables $u,v$ (\ref{uv}), equations (\ref{g1a})-(\ref{g2a}) become
\begin{align} \label{g1b}
\eta_{t} + u_{x} +v_{y} & + \alpha  \left(\eta_{x} u +\eta_{y} v
+\eta(u_{x}+v_{y})\right) \\ & -\frac{1}{6}  \epsilon 
   \left(u_{xxx} + 2u_{xyy}+v_{yyy} \right)    = 0, \nonumber \\   \label{g2b}
\eta_{x} + u_{t} 
& + \alpha  \left(u u_{x}+v v_{x} \right) 
 -\frac{1}{2} \epsilon \left(u_{yyt} +u_{xxt}  \right) 
  =0.\end{align}  
Equations (\ref{g1a}) and consequently (\ref{g1b}) differ from equations (\ref{g1F}) and (\ref{g2F}) obtained from incorrect velocity potential (\ref{F11}) by the factor in front of the term $f_{xxyy}$.

Let us follow the authors' approach with the correct equations (\ref{g1b})-(\ref{g2b}). In zeroth order we have  (\ref{b0ord}). In first order assume the correction functions in the same form  (\ref{c1aF})-(\ref{c1bF}).

Substitution to (\ref{g1b}) and limitation to first order terms gives
\begin{equation} \label{BxCx}
 \eta_{t} +\eta_{x}+ \alpha\left(a\,\eta_{y}+B_{x} +2\eta\eta_{x}\right)+ \varepsilon\!\left(\!C_{x}-\frac{1}{6}\eta_{3x}-\frac{1}{3}\eta_{x2y}\!\right)=0. 
\end{equation}
Substitution to  (\ref{g2b}) and retention of terms up to first order gives 
\begin{equation} \label{BtCt}
\eta_{t} +\eta_{x}+ \alpha\left(B_{t}+\eta\eta_{x}\right)+ \varepsilon\left(C_{t}-\frac{1}{2}\eta_{2yt}-\frac{1}{2}\eta_{2xt}\right)=0.
\end{equation}
Substraction of (\ref{BtCt}) from (\ref{BxCx}) gives
\begin{align} \label{1corr}
\alpha &(B_x-B_t+\eta\eta_x+a\eta_y)\\  +\varepsilon&(C_x-C_t
-\frac{1}{6}\eta_{3x}-\frac{1}{3}\eta_{x2y}+\frac{1}{2}\eta_{2yt}+\frac{1}{2}\eta_{2xt})=0 .\nonumber
\end{align}
In (\ref{1corr}), we use the properties 
$B_{t}=-B_{x}, ~C_{t}=-C_{x}, ~\eta_{2yt}=-\eta_{2yx}, ~\eta_{2xt}=-\eta_{3x}$ valid in zeroth order. Since expressions in (\ref{1corr}) are already in first order it is sufficient. Recall that the general form, e.g., $B_t=-B_x +\alpha Ba +\varepsilon Be$, and so on, do not change the further results since after insertion into (\ref{1corr}) second order terms have to be rejected.
These properties and freedom of $\alpha,\varepsilon$ allow us to obtain
 simple differential equations for $B$ and $C$ in the form
 $$ 2B_x +\eta\eta_x+a\eta_y=0, \quad 2C_{x}-\frac{2}{3}\eta_{3x} - \frac{5}{6}\eta_{xyy}=0. $$
Integrating above equations  one obtains 
\begin{equation} \label{BC}
 B = -\frac{1}{4} \eta^{2} -\frac{1}{2} a\! \int\! \eta_{y}\,dx, \quad \mbox{and}\quad C= \frac{1}{3}\eta_{xx} + \frac{5}{12}\eta_{yy}. 
\end{equation}
Then, the function $u$ becomes 
\begin{equation} \label{u1ord}
u = \eta +\alpha\!\left(\!  -\frac{1}{4} \eta^{2} -\frac{1}{2} a\!\int\! \eta_{y}\,dx\! \right) +\varepsilon\left(\! \frac{1}{3}\eta_{xx} + \frac{5}{12}\eta_{yy} \!\right).
\end{equation}
With $u$ in the form (\ref{u1ord}), both Boussinesq's equations should reduce to the same wave equation. Indeed, this is the case, and the final first order wave equation receives the following form
\begin{equation} \label{b1nasze}
\eta_{t}+\eta_{x}+\alpha\left(\frac{3}{2}\eta\eta_{x}+\frac{1}{2} a\,\eta_{y}\right)+ \varepsilon\left(\frac{1}{6}\eta_{xxx}+ \frac{1}{12}\eta_{xyy}\right)=0.
\end{equation}

Although the steps from assumed form of first order functions (\ref{c1aF})-(\ref{c1bF}) to equations (\ref{Fok26}) \cite[Eq.~(26)]{Fok} or (\ref{b1nasze}) are mathematically correct, the final equations are incorrect because the assumptions (\ref{c1aF})-(\ref{c1bF}) violate the properties of the velocity potential. Deatils ere explained in the next section.

\section{Critics}\label{s4}

The authors 
treat $u,v$ as independent functions, ignoring that, in fact, $u,v$  
are partial derivatives of the same function $f$. Since $u=f_{x}$ and $v=f_{y}$, and these functions and their partial derivatives should be continuous, then the fundamental condition has to be fulfilled
\begin{equation} \label{NesCon}
u_{y}=v_{x}\equiv f_{xy}. \end{equation} 
This condition is not fulfilled by the functions $u,v$ in equations (\ref{c1aF})-(\ref{c1bF}) which are necessary for the authors to derive new (2+1)-dimensional equation.   The requirement that condition (\ref{NesCon}) is fulfilled by  
(\ref{c1aF})-(\ref{c1bF}) leads from one side to 
$$f_{xy}\equiv u_{y}=\eta_{y} +\alpha B_{y}+\varepsilon C_{y} $$
and from the other side to
$$f_{xy}\equiv v_{x}= 0.$$ 
So, the condition ~$u_{y}=v_{x}$~ (\ref{NesCon})~ can be fulfilled only when 
$\eta_{y} +\alpha B_{y}+\varepsilon C_{y}=0$~ which, due to freedom of $\alpha,\varepsilon$, imply  ~$B_{y}=C_{y}=\eta_{y}=0$ !  This means that the surface profile does not dependent on $y$ (there is translation symmetry in $y$-coordinate).
In other words, searching for the solution in the form (\ref{c1aF})-(\ref{c1bF}) {\bf reduces the problem to be only (1+1)-dimensional} (with $\eta_{y}=0$ the final equation reduces to the usual KdV equation).
Therefore, the derived first order (2+1)-dimensional evolution equation cannot be the solution to the Boussinesq set (\ref{g1F})-(\ref{g2F}). In the same way the equation (\ref{b1nasze})  cannot be the solution to the correct Boussinesq's set  (\ref{g1a})-(\ref{g2a}).

Can we extend the form of corrections (\ref{c1aF})-(\ref{c1bF}) in such a way that they fulfill the condition (\ref{NesCon})? We assume
the most general form of first order functions $u,v$, that is
\begin{eqnarray} \label{w1}
u &=& \eta +\alpha\, B(x,y,t) +\varepsilon\, C(x,y,t), \\ \label{w2}
v &=&  a(x,t) +\alpha\, R(x,y,t) +\varepsilon\, S(x,y,t).
\end{eqnarray}
Inserting (\ref{w1})-(\ref{w2})  into (\ref{g1b})-(\ref{g2b}) and neglecting terms of second order one obtains
\begin{align} \label{Bw1}
 \eta_{t} +\eta_{x} & + \alpha\left( a\,\eta_{y}+B_{x} +2\eta\eta_{x}+R_{y}\right) \\ & + \varepsilon\left(C_{x}-\frac{1}{6}\eta_{xxx}-\frac{1}{6}\eta_{xyy}+S_{y}\right)=0 \nonumber \\
  & \mbox{and} \nonumber \\ \label{Bw2}
\eta_{t} +\eta_{x} & + \alpha\left(B_{t}+\eta\eta_{x}+a\,a_{x}\right)+ \varepsilon\left(C_{t}-\frac{1}{2}\eta_{yyt}-\frac{1}{2}\eta_{xxt}\right)=0.
\end{align} 
Substraction of (\ref{Bw2}) from (\ref{Bw1}) and use the properties \\
$B_{t}=-B_{x},~ ~C_{t}=-C_{x}$ allows us to obtain two equations
\begin{align} \label{rB}
2B_{x} & +\eta\,\eta_{x} -a\,a_{x}+a\,\eta_{y} +R_{y}=0, \quad \mbox{and} \\ \label{rC}
2C_{x} & -\frac{2}{3}\left(\eta_{xxx}+\eta_{xyy} \right)+S_{y}=0.
\end{align}
Integration over $x$ gives
\begin{align} \label{B&C}
 B  & = -\frac{1}{4} \eta^{2}+\frac{1}{4} a^{2} -\frac{1}{2} a\! \int\! \eta_{y}\,dx -\frac{1}{2}\! \int\! R_{y}\,dx, \\  C & = \frac{1}{3}(\eta_{xx} +\eta_{yy})-\frac{1}{2}\! \int\! S_{y}\,dx.  \nonumber
\end{align}
So, in principle, the functions 
\begin{align} \label{W1}
u &= \eta +\alpha \left( -\frac{1}{4} \eta^{2}+\frac{1}{4} a^{2} -\frac{1}{2} a\! \int\! \eta_{y}\,dx -\frac{1}{2}\! \int\! R_{y}\,dx  \right) \nonumber \\
&  \hspace{4.5ex} +\varepsilon \left(\frac{1}{3}(\eta_{xx} +\eta_{yy})-\frac{1}{2}\! \int\! S_{y}\,dx \right),  \quad \mbox{and} \\ \label{W2}
v &=  a(x,t) +\alpha\, R(x,y,t) +\varepsilon\, S(x,y,t),
\end{align}
can make Boussinesq's equations compatible. But, is it possibile to find such functions $a,R,S$ which ensure fulfilling the condition (\ref{NesCon}), that is ensure $u_{y}=v_{x}$?
Because 
\begin{align} \label{W1a}
u_{y} &= \eta_{y} +\alpha \left(-\frac{1}{2} \eta\eta_{y}-\frac{1}{2} a\! \int\! \eta_{yy}\,dx-\frac{1}{2}\! \int\! R_{yy}\,dx  \right) \nonumber \\ & \hspace{4.5ex} +\varepsilon \left(\frac{1}{3}(\eta_{xxy} +\eta_{yyy})-\frac{1}{2}\! \int\! S_{yy}\,dx \right) \hspace{5ex}\\ \label{W2}
v_{x} &= a_{x}+\alpha\, R_{x} +\varepsilon\, S_{x},
\end{align}
this task seem to be hopeless.

\section{Another possibile approach}

The above considerations imply that in a shallow water problem, it is practically impossible to obtain (2+1)-dimensional evolution equation for the profile of surface wave $\eta(x,y,t)$, even in first order approximation. 
However, this model can supply differential equations of any order for the function $f(x,y,t)\equiv \phi^{(0)}(x,y,t)$. Below, we present such an equation in first order. 

Recall, that from (\ref{G2}) and the velocity potential (\ref{pot8}) one obtains the first Boussinesq's equation in the form (\ref{g1a}).  The corresponding  Boussinesq's equation resulting from (\ref{G3}) is 
\begin{equation} \label{e46}
\eta + f_{t}+\frac{1}{2}\alpha\left(f_{x}^{2}+f_{y}^{2}\right) -\frac{1}{2} \varepsilon \left(f_{xxt}+f_{yyt} \right) =0.
\end{equation}
Now, inserting $\eta=-\left( f_{t}+\frac{1}{2}\alpha\left(f_{x}^{2}+f_{y}^{2}\right) -\frac{1}{2} \varepsilon \left(f_{xxt}+f_{yyt} \right)\right)$ from (\ref{e46}) into (\ref{g1a}) and neglecting second order terms one obtains first order evolution equation for the function $f(x,y,t)$ 
\begin{align} \label{KP1ord}
f_{xx} & +f_{yy} -f_{tt} + \alpha \left[ -f_t \left(f_{xx} +f_{yy}  \right)- 2\left(f_x f_{xt}+f_y f_{yt} \right)\right] \\ & +\varepsilon \left[ \frac{1}{2}\left(f_{xxtt} +f_{yytt}\right) -\frac{1}{6} \left(f_{xxxx} +2 f_{xxyy}+f_{yyyy} \right) \right]  =0.\nonumber 
\end{align}
It can be further simplified utilizing zeroth-order solution $f_{tt}=f_{xx}+f_{yy}$. Since the term $\varepsilon \frac{1}{2}\left(f_{xxtt} +f_{yytt}\right)$ is already first order, we can use the equalities $$ f_{xxtt}=(f_{tt})_{xx}=(f_{xx}+f_{yy})_{xx}=f_{xxxx}+f_{xxyy},$$  $$ f_{yytt}=(f_{tt})_{yy}=(f_{xx}+f_{yy})_{yy}=f_{xxyy}+f_{yyyy}$$ 
to replace it by
$ \varepsilon \frac{1}{2}\left(f_{xxxx}+2f_{xxyy}+f_{yyyy}\right).$
Therefore an equivalent, simpler form of (\ref{KP1ord}), still valid up to first order, is
\begin{align} \label{KP1ordS}
f_{xx} & +f_{yy} -f_{tt} + \alpha \left[ -f_t \left(f_{xx} +f_{yy}  \right)- 2\left(f_x f_{xt}+f_y f_{yt} \right)\right] \\ & +\varepsilon \left(\frac{1}{3} \left(f_{xxxx} +2 f_{xxyy}+f_{yyyy} \right) \right)  =0.\nonumber 
\end{align}
If the solution to (\ref{KP1ord}) or  (\ref{KP1ordS}) is known, the equation (\ref{e46}) supplies the surface profile function $\eta(x,y,t)$.

In general, for a flat bottom, one can extend this procedure to arbitrary order. However, even in first-order approximation partial differential equation (\ref{KP1ordS}) corresponding to (2+1)-dimensional shallow water problem is highly complicated. There is little hope to find either a solution to (\ref{KP1ordS}) or a (2+1)-dimensional wave equation for wave profile $\eta(x,y,t)$ without further significant simplifications.

\section{Conclusions}
We have proved that the (2+1)-dimensional KdV-type equation \cite[Eq.~(26)]{Fok} has been inconsistently obtained by the authors and therefore cannot describe (2+1)-dimensional surface waves. Moreover, we have shown that when assumptions for first order functions \cite[Eqs.~(14)-(15)]{Fok} are used consistently with the properties of the velocity potential, then the solution reduces to the usual KdV equation. Additionally, we have demonstrated that even a consistent extension of the authors' method \cite{Fok} gives no hope for obtaining appropriate first order (2+1)-dimensional evolution equation for shallow water problem.



\end{document}